# Modelling uncertainty in coupled electricity and gas systems—is it worth the effort?

Iegor Riepin[*,1], Thomas Möbius[*] and Felix Müsgens[*]

[*]Brandenburg University of Technology Cottbus-Senftenberg (BTU)

## Abstract

The interdependence of electricity and natural gas markets is becoming a major topic in energy research. Integrated energy models are used to assist decision-making for businesses and policymakers addressing challenges of energy transition and climate change. The analysis of complex energy systems requires large-scale models, which are based on extensive databases, intertemporal dynamics and a multitude of decision variables. Integrating such energy system models results in increased system complexity. This complexity poses a challenge for energy modellers to address multiple uncertainties that affect both markets. Stochastic optimisation approaches enable an adequate consideration of uncertainties in investment and operation planning; however, stochastic modelling of integrated large-scale energy systems further scales the level of complexity. In this paper, we combine integrated and stochastic optimisation problems and parametrise our model for European electricity and gas markets. We analyse and compare the impact of uncertain input parameters, such as gas and electricity demand, renewable energy capacities and fuel and $CO_2$ prices, on the quality of the solution obtained in the integrated optimisation problem. Our results quantify the value of encoding uncertainty as a part of a model. While the methodological contribution should be of interest for energy modellers, our findings are relevant for industry experts and stakeholders with an empirical interest in the European energy system.

[1]Corresponding author: Iegor Riepin, BTU Cottbus-Senftenberg, Siemens-Halske-Ring 13, 03046, Cottbus, Germany. E-mail: iegor.riepin@b-tu.de. This work is supported by the postgraduate scholarship (GradV).

# 1 INTRODUCTION

Modelling energy markets is immensely important for academic purposes, business decision-making and governmental projections to address energy transitions and climate change. Modern energy models entail extensive databases, intertemporal dynamics and a multitude of decision variables, all of which are necessary to capture the complex behaviour of energy markets.

Ongoing energy transitions present new challenges for energy modellers by increasing the interconnection of energy markets. These challenges require consideration of multiple interdependencies between the models of different sectors, such as the electricity and gas sectors (ENTSOs, 2019). The growing role of gas-fired plants in renewable-based electricity markets and the increasing dependence on natural gas imports make this issue particularly important for the European energy modelling community. Thus, the number of model-based studies focusing on integrated modelling of electricity and gas markets has grown in recent years. The advantages of integrated modelling come at a cost in the form of higher computational complexity stemming from simultaneous optimisation of both markets with their own intertemporal dynamics and relevant infrastructure.

Another factor troubling energy modellers is the uncertainties of input data that characterise the future development of electricity and gas sectors. These include primary energy carrier prices, technological developments (e.g. installed capacities of subsidised renewable generation), regulations and political context. Most studies focusing on integrated optimization of electricity and gas markets to date have been using a deterministic approach. This is likely due to the fact that stochastic models—even for a single sector—are complicated to construct and expensive to run. In many cases, however, ignoring uncertainty leads to poor decisions (see: Egging, 2010; van der Weijde & Hobbs, 2012).

Therefore, those working on state-of-the-art energy models must decide whether to incorporate uncertainty into large-scale integrated models and scale computational complexity even further or to neglect uncertainty and accept a suboptimal set of decisions. When addressing this trade-off, a natural question to ask is whether neglecting uncertainty results in nearly optimal or in an inaccurate solution.

The objective of this paper is to address the modellers' trade-off with a quantitative analysis. We evaluate how much difference it makes to the quality of the decisions reached (in our application – the quality of investment decision in power generation mix) when uncertainty is explicitly encoded (or removed) as a part of a model.

Methodologically, this paper contributes to the modelling literature in two major ways. First, we combine integrated and stochastic optimisation models for electricity and gas markets. On the one hand, a stochastic two-stage optimisation framework captures the multistage nature of capacity planning under uncertainty; we contribute to the literature stream on long-term investment problems in integrated energy systems with consideration of parametric uncertainty. On the other hand, an integrated framework allows us to extend the single market uncertainty evaluations by tracing the effects of uncertainty across the integrated energy system. Second, we investigate and compare the isolated effects of parametric uncertainty. We present a framework that analyses a wide range of uncertain model inputs to quantify and compare the effects. Furthermore, we provide in-depth insight into how parametric uncertainty affects residual loads and variable costs of power generation.

We apply our model to a stylised representation of the European market. The majority of model-based studies on Europe rely on ENTSOs data for the future development of the electricity and gas sectors. We parametrise our model the same way by basing our data primarily on the TYNDP report from ENTSOs (2018). Thus, we believe this research is a useful reference for energy modellers and industry experts. Our discussion focuses on the concept of the expected costs of ignoring uncertainty, which provides a solid theoretical background to understand the modellers' trade-off (Birge & Louveaux 2011). In order to facilitate the transparency and reproducibility of our results, as well as to encourage further research on model coupling and uncertainty, our model code, associated data and scripts for result processing are all published on the public GitHub repository.[2]

# 2 INTEGRATED AND STOCHASTIC MODELLING OF ELECTRICITY AND GAS MARKETS IN THE LITERATURE

## 2.1 Integrated modelling

As already stated, interest in the representation of complex interdependencies between the electricity and gas sectors in energy models has grown in recent years. The early studies on the interplay between the two sectors were limited in their application to (i) simplified or no representations of intertemporal dynamics, (ii) toy parametrisation and (iii) soft- or hard-linking approaches.[3] For example, an optimisation model on the natural gas and electricity sectors over a single time period was discussed by An et al. (2003) and Geidl & Andersson (2007). A hard-linking approach is used by Ohishi & de Mello (2005), among others. These authors highlight the importance of integrated modelling for economic and secure energy system operation based on a toy system model. Bartels & Seeliger (2005) use a hard-linking approach to analyse the long-term impact of $CO_2$-emission trading on the two markets with a pan–European geographical scope. More examples are provided by Rubio-Barros et al. (2008), who present a literature survey on integrated natural gas and electricity system planning literature and highlight economic and market-related aspects.

Advancements in computing power and mathematical models paired with the challenges of the energy transition process have facilitated the development of more sophisticated models. Such models address the limitations of earlier research and, thus, are characterised by (i) a larger time scope and higher time resolution with complex intertemporal dynamics, (ii) parametrisation to a regional or pan–European geographical scope and (iii) integrated (simultaneous) optimisation of both sectors. Chaudry et al. (2008) investigate the importance of gas storage in the context of integrated system stability based on the British gas and electricity network. Möst & Perlwitz (2009) focus on European gas supply prospects through 2020 and their relevance for the power sector in the context of emission trading. Lienert & Lochner (2012) focus on the short- and long-term dynamics between natural gas and electricity markets using the pan–European integrated model; they highlight that quantitative models that do not consider interdependencies between the two markets produce results with systematic deviations from a more realistic integrated optimisation. Abrell & Weigt (2016b) run several long-term gas market scenarios to capture their impact on power plant investments and short-term supply shock scenarios to analyse spatial feedback towards the electricity system. Deane et al. (2017) construct an integrated model with a daily temporal resolution to examine the impact of gas supply interruptions on power system operation and gas flow in the European market. Several authors focus on coordinated expansion planning problems and investigate the potential for substitution effects between investments in generation and transmission across both sectors (Abrell & Weigt, 2016a; Zhao et al., 2018; Chaudry et al., 2014). Ameli et al. (2017) and Clegg & Mancarella (2016) discuss the value of gas network infrastructure flexibility in supporting the cost-effective operation of power systems.

---

[2] See: github.com/Irieo/IntEG
[3] *Soft-linking* is defined as a model coupling approach in which information is processed and transferred manually by the modeller; *hard-linking* (usually associated with an iterative solution approach) is defined as a model coupling approach in which input/output information transfer is handled by an algorithm (Helgesen, 2013).

## 2.2 Stochastic modelling

Actors in modern liberalised electricity markets face multiple uncertainties. These are driven by the development of prices for primary energy carriers, the structural changes in the energy sector (e.g. introduction of carbon markets or nuclear phase-outs in several countries around the globe) and regulation (subsidization of renewable generation, decarbonisation policy, the introduction of the carbon price floor in the UK, etc). There are also uncertainties in natural gas markets, where many factors on the demand side (weather, decarbonisation policy) and the supply side (gas reserves) are inherently uncertain. Gas markets are also subject to structural breaks, such as recent drops in production capacity in North-Western Europe or US shale gas revolution.

It is important to understand the impact of uncertain input parameters on the model outputs, conclusions or policy recommendations. Stochastic programming, first conceived by Dantzig (1955) as a framework for decision-making under uncertainty, has been successfully applied to energy models for decades. Möst & Keles (2010) provide a comprehensive survey of stochastic modelling approaches for liberalised electricity markets while Egging (2010) provides a structured overview of stochastic market models and algorithms for both sectors; both of these reviews show that stochastic modelling of energy markets can take on various mathematical forms with different purposes and solution algorithms. In this section, however, we focus primarily on applications of multi-stage optimisation models for short- and mid-term generation and long-term system expansion planning, which are most relevant for this paper.

Regarding electricity markets, Müsgens & Neuhoff (2006) use stochastic optimisation to analyse the impact of the daily wind feed-in on dispatch decisions and the value of updating wind forecasts. Fürsch et al. (2012) use a multi-stage stochastic programming approach to optimise power plant investments along uncertain renewable energy development paths. Van

der Weijde & Hobbs (2012) use two-stage stochastic optimisation for electricity grid reinforcement planning under uncertainty; they highlight the fact that ignoring risk in transmission planning for renewables has quantifiable economic consequences while considering uncertainty yields decisions with expected costs lower than those from traditional deterministic planning methods. Seljom & Tomasgard (2015), studying the impact of stochastic wind feed-in on optimal generation capacity, conclude that the stochastic representation of intermittent renewables in long-term investment models provides more reliable results for decision-makers. Möbius & Müsgens (2017) study the impact of uncertain wind feed-in on long-term market equilibria. Schwarz et al. (2018) present a two-stage stochastic problem for optimising investment and operation decisions in a decentralised energy system.

Regarding gas markets, Zhuang (2005) and Zhuang & Gabriel (2008) develop an extensive-form stochastic complementarity problem and provide a small-scale natural gas market implementation. Egging (2010) develops a stochastic multi-period global gas market model. The author concludes that stochastic modelling shows hedging behaviour that affects the timing and magnitude of capacity expansions, significantly affecting local market situations and prices. Fodstad et al. (2016) analyse the impact of uncertainty about future European natural gas consumption on optimal investments in gas transport infrastructure and conclude that the option value of delaying investments in natural gas infrastructure until more information is available in 2020 is very limited due to the low costs of overcapacity. They also find, however, structural differences between the infrastructure investments derived from the stochastic model and those from the deterministic model.

## 2.3 Identifying the research gap

Our literature review reveals two important research gaps. First, few studies to date have combined market integration and stochastic elements. These are however, comprise toy energy systems (Ordoudis et al., 2019; Su et al., 2015), focus on a single uncertainty (Riepin et al., 2018; Zhao et al., 2017) or are static (Zhao et al., 2018). This observation is bolstered by Deane et al. (2017), who state that while uncertainty in electricity and gas systems are not a recent phenomenon, their impacts on integrated systems are not well examined. Second, no study has yet to provide a framework to ignore isolated uncertain model inputs and subsequently quantify and compare their effects. Thus, there is still no systematic understanding of which isolated parametric uncertainty most substantially affects long-term planning decisions.

This paper serves to fill in these two research gaps. We present an applied methodology and an open-source model to examine the trade-off between complexity from integrated optimisation of gas and electricity systems and a wide range of parametric uncertainties.

# 3 METHODOLOGY

We construct and apply an integrated stochastic bottom-up optimisation model for European electricity and gas markets. The model's objective function minimises the total costs, comprising expected discounted capital and operating costs for both sectors. The optimal solution implies that all arbitrage opportunities across time and space are exhausted to the extent that the infrastructure constraints of the integrated system permit. The results of the model include spatial and temporal decisions on both investment in power generation units and the production, transportation and storage of electricity and gas. The incorporation of storage and investment decisions into the optimisation problem requires (and enables) intertemporal optimisation. We introduce one-stage deterministic optimisation and two-stage stochastic optimisation approaches in Section 3.1. Once all model runs are complete, we analyse the results to determine whether neglecting uncertainty results in an inaccurate solution. The theoretical background for this approach is detailed in Section 3.2.

The gas market components in our model (input data and decision variables) have a temporal resolution of 12 consecutive months while the electricity market components have one of 350 representative hours for each modelled year. The temporal scope is 2020–2030. Model simulations are performed for three representative years: 2020, 2025 and 2030. This allows us to capture both the short-term market operations and long-term investments dynamics.

The model structure is made up by a network of nodes. A node represents a country or a group of several countries from one region. Nodes are connected by electricity and gas transmission infrastructure. The geographical scope covers most European member states, Norway, Switzerland, the United Kingdom and several non-European major gas exporters (Russia, the United States, Algeria, Libya, Nigeria and Qatar). We provide a full list of the countries considered in our model in Appendix C.

The markets are combined via fuel linkage; both the gas demand for the power sector and the price for gas-fired electricity generation are modelled endogenously. Fluctuations in natural gas demand from the power and non-power sectors induce gas price volatility. Thus, our modelling approach ensures endogenously defined spatial and temporal gas price patterns; as gas price is the cost input for gas-fired units, its volatility affects dispatch and investment decisions in the electricity sector.

Integrated modelling has several advantages over soft- and hard-linking approaches. First, it excludes the convergence criteria used in iterative linking approaches. These criteria are usually based on the rate of change between model outputs over subsequent iterations. Large-scale energy models solved iteratively may encounter regular convergence problems (Holz et al., 2016; Helgesen, 2013). Second, integrated optimisation of two sectors ensures that the optimal solution includes reliable marginal cost estimators. Note that marginal electricity generation costs are derived from the dual variables of each node's energy balance constraints. Thus, the optimal solution ensures that a relaxation of these constraints (by one MWh of gas or electricity) returns true marginal savings from producing, transporting and storing that energy unit.[4] Third, integrated modelling allows us to conveniently handle a large number of model runs, which are necessary to answer this paper's research questions (see discussion of scenarios and modes in Section 3.3).

We provide a complete model formulation in Section 3.4. All of the data used are from publicly available sources. In section 3.5, we discuss our assumptions on both sectors' demand and supply structure and transmission infrastructure. The model is formulated in GAMS[5] and solved with a CPLEX solver.

---

[4] Thus, marginal costs can be considered as price indicators in a competitive market.

## 3.1 Optimization approaches

In this section, we introduce the one-stage deterministic optimisation and the two-stage stochastic optimisation approaches.

Consider a linear optimization problem (LP), where $x$ represents a vector of variables, $c$ and $b$ are parameter vectors (i.e. known coefficients), $A$ is a matrix of parameters, and $T$ denotes a matrix transpose. The inequalities are the constraints that specify a convex polytope over which the objective function is to be optimised. The optimal deterministic solution (DS) is to find a vector $x$ that minimizes the objective under the set of relevant constraints:

$$DS \coloneqq \min_{x} c^T x$$
$$s.t. Ax \geq b, x \geq 0 \quad (1)$$

Due to their relative simplicity, these models can be solved with a high degree of empirical detail. Hence, they are a widely used 'work-horse' in energy system modelling.

However, energy system forecasters face multiple uncertainties, such as primary energy carrier prices, technological developments, regulations and political context. The optimisation problem and the resulting decisions depend on these uncertainties. They are particularly relevant when analysing investment decisions in energy systems, which are largely irreversible and involve a high share of total generation costs. An approach to explicitly incorporate uncertainty is to represent the multi-stage nature of investment planning in a two-stage stochastic model. It implies that optimal *first-stage* investment decisions in power generation technologies must be made before the information on uncertain factors is revealed; while *second-stage* dispatch decisions are made after uncertainty is revealed.

The classical two-stage stochastic linear problem (SP) can be formulated as follows (Birge & Louveaux, 2011):

$$\min_{x} c^T x + E_{\omega}[Q(x, \omega)]$$
$$s.t. Ax = b, x \geq 0 \quad (2)$$

where

$$Q(x, \omega) \coloneqq \min_{x} q_{\omega}^T y$$
$$s.t. T_{\omega} x + W_{\omega} y \geq d_{\omega}, y \geq 0 \quad (3)$$

Where $x$ represents the vector of first-stage variables, y is the vector of second-stage variables, and $\omega$ is the vector of uncertain data for the second stage (i.e. the vector of possible scenarios). The parameter $c^T$, the matrix $A$, and the right-hand-side vector b of the first stage are assumed to be known with certainty. Problem (2) seeks a first-stage decision to minimise the costs that occur at the first stage and the *expected* costs of second-stage (recourse) decisions. Problem (3) seeks second-stage decisions that minimise the second-stage costs. Second-stage decisions are restricted by the first-stage decisions of x, the matrix $T$, the matrix $W$, and the right-hand-side vector $d$. Note that the parameters $(q, T, W, d)$ are actual realisations of uncertain data.

---

[5] General Algebraic Modeling System, see: https://www.gams.com/.

By solving a stochastic problem, we obtain an optimal solution $\bar{x}$ of the first-stage problem and optimal solutions $\bar{y}_n$ of the second-stage problem for each realisation of $\omega_n$. Given $\bar{x}$, each $\bar{y}_n$ corresponds to an optimal second-stage decision corresponding to a realisation of the respective scenario. In the context of this paper, the solution of a stochastic problem (in the sense of minimising total expected costs) defines (i) the optimal electricity generation investment (which must hold for all scenarios) and (ii) scenario-dependent optimal dispatch decisions of all assets for both electricity and gas components.

## 3.2 The expected cost of ignoring uncertainty

Stochastic problems are often avoided in practice because they are computationally difficult to solve. Many real-world problems are addressed with simpler approaches. For example, one can solve several deterministic programs—each corresponding to one particular scenario—and then combine solutions using a heuristic rule. The approximation problem most often discussed in the context of two-stage stochastic problems is the expected value problem (EVP), a problem wherein the uncertain parameters are replaced by their expected values. Consider a two-stage stochastic problem as formulated in section 3.1; a scholastic solution (SS) is defined as follows:

$$SS := \min_{x} c^T x + E_\omega[Q(x, \omega)] \tag{4}$$

The EVP is constructed by setting $\bar{\omega} = E_\omega \omega$. Thus, a solution of an EVP is:

$$EVS := \min_{x} c^T x + Q(x, \bar{\omega}) \tag{5}$$

Figure 1 illustrates the modelling of an SP versus that of an EVP.

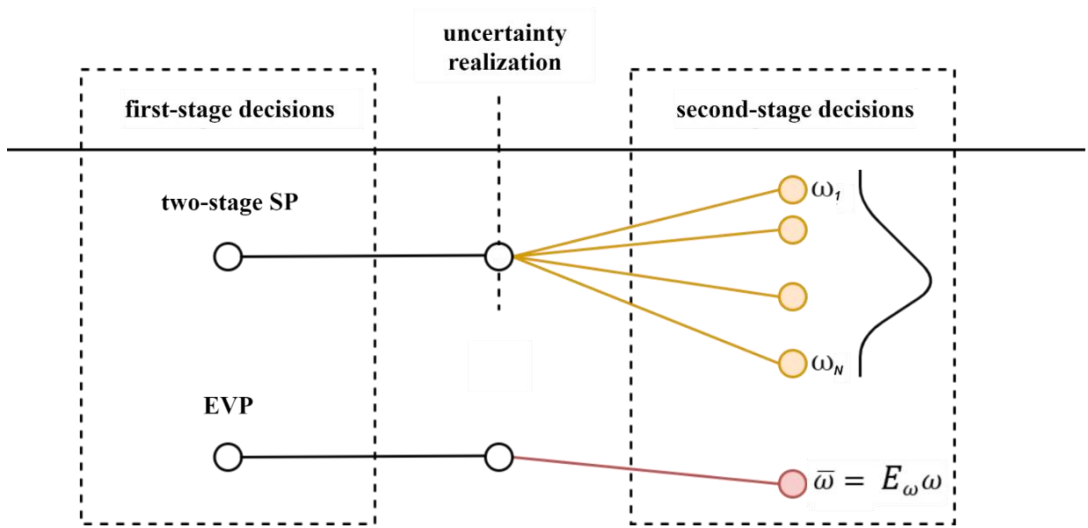

Figure 1: A two-stage stochastic problem and an approximation (EVP).

As we have already pointed out, problem (5) is an approximation problem, meaning it does not consider uncertainty. An alternative is to explicitly encode uncertainty into the model by setting a probability distribution for uncertain parameters, as shown in problems (2) and (3). It is crucial for us to understand whether ignoring uncertainty reduces the quality of the decisions reached. The theoretical answer to this is given in the literature by the concept of the expected costs of ignoring uncertainty (see: Birge & Louveaux, 2011; Birge, 1982).

To reach this understanding, let $\bar{x}(\bar{\omega})$ denote an optimal first-stage decision in an EVP. Then, constraining the stochastic problem with the first-stage decisions of the EVP shows how well a decision $\bar{x}(\bar{\omega})$ performs. The expected result of imposing $\bar{x}(\bar{\omega})$ into a stochastic problem is denoted by EEV:

$$EEV := E_\omega[\varphi(\bar{x}(\bar{\omega}), \omega)] \qquad (6)$$

The expected cost of ignoring uncertainty is defined as:

$$ECIU = EEV - SS \qquad (7)$$

The ECIU is useful because it describes the value of considering the full range of uncertainties in a stochastic model rather than that of using a deterministic problem. Thus, the metric can be interpreted as *the expected cost of assuming that the future is certain*.

### 3.3 Scenario composition

*Introducing scenarios*

This paper quantifies the ECIU relative to the EVP discussed above based on three scenarios from TYNDP (presented in Section 3.5). Furthermore, we compute and discuss the ECIUs when each of ENTSOs' scenarios are chosen as the reference for a deterministic model. This is done by replacing the EVP from problem (5) with the data for a specific scenario, $\bar{\omega} = \omega_n$. Throughout the discussion, we use the term '*naïve problem*' (NP) to refer to the set of four (each $\omega_n$ and $\bar{\omega}$) possible deterministic problems chosen by a system planner when uncertainty in data is ignored (albeit present).

Thus, the solution to an NP (NPS) is:

$$NP: NPS = \min_x c^T x + Q(x, \hat{\omega} \in \{\omega_n, \bar{\omega}\}) \qquad (8)$$

Consequently, we incorporate $\bar{x}(\omega_n)$ into a stochastic problem to evaluate the expected result of each NPS:

$$EEV_n = E_\omega[\varphi(\hat{x}(\hat{\omega}), \omega)] \qquad (9)$$

This allows us to compute the scenario-specific ECIUs:

$$ECIU_n = EEV_n - SS \qquad (10)$$

*Introducing stochastic model modes*

We define two stochastic model modes for this study: '*all parameters*' and '*isolated parameters*'. In the first mode, a vector of uncertain data in a stochastic problem includes all five uncertain parameters. For example, scenario branch EUCO includes gas demand, electricity demand, installed RES capacity, fuel prices and $CO_2$ prices, which all have a path as defined in the EUCO scenario of the TYNDP. The same applies to the other two branches of a stochastic problem. Thus, optimal first-stage investment decisions in power generation technologies must be made with consideration for a composite uncertainty.

In the second mode, we provide more in-depth analysis by isolating the effects of parametric uncertainty. In this mode, a vector of uncertain data in a stochastic problem includes a single parameter. Intuitively, the development paths of the other four parameters (which are known in this set-up) have their own effects, which we capture by computing a matrix consisting of combinations of isolated parametric uncertainties and the possible development paths of known parameters.

*Combining scenarios and modes*

The composition of modes and scenarios is illustrated in Figure 2. Throughout the discussion of our results, the term '*scenario branch*' refers to the branches of a stochastic model (EUCO, ST, DG) that define possible development paths of uncertain parameters; the term '*scenario*' refers to (i) the system planner's choice for solving a deterministic problem (in 'all parameters' mode) and (ii) the development paths for known parameters (in 'isolated parameters' mode).

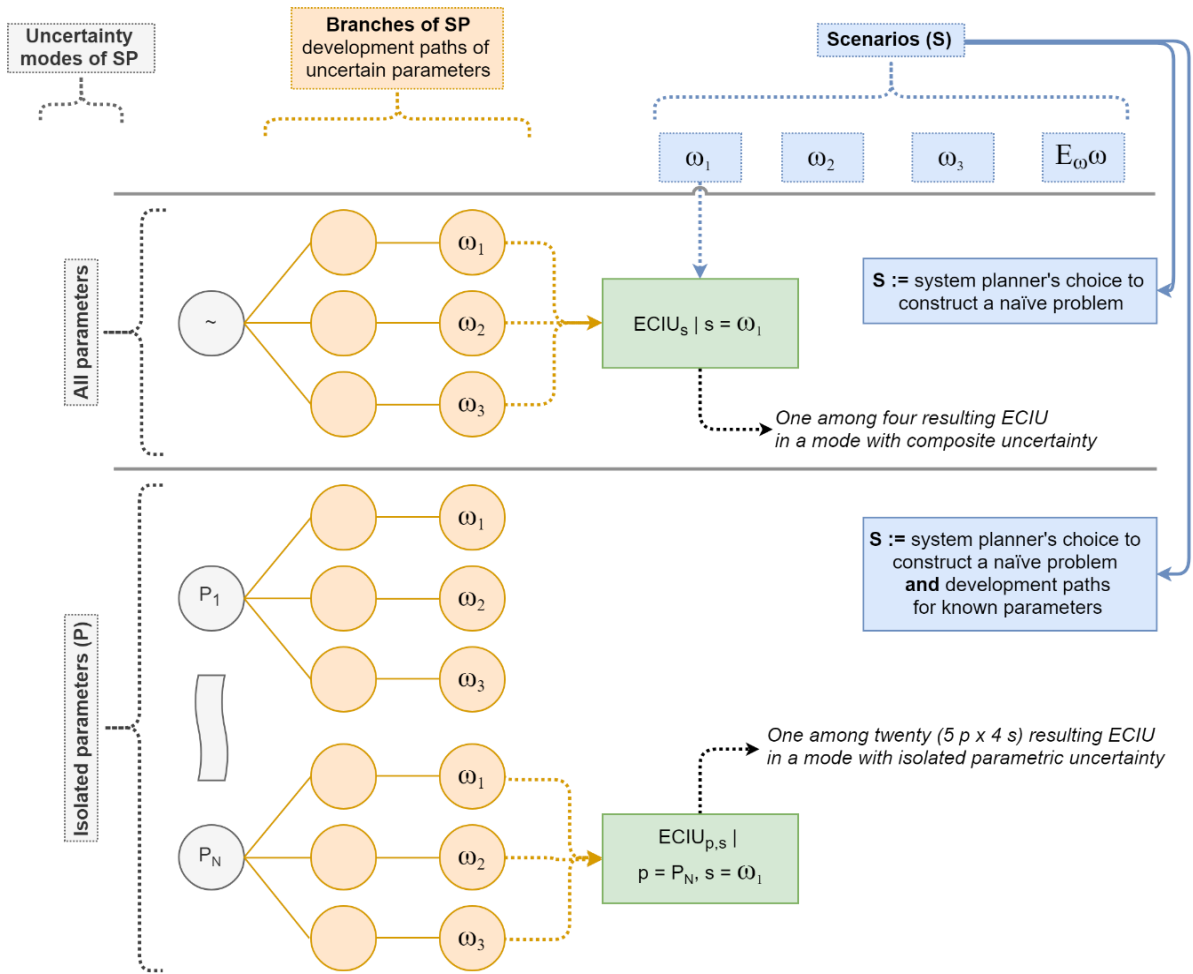

Figure 2: Graphical illustration of scenario composition.

## 3.4 Model formulation

**Nomenclature**

| Abbreviation | Dimension | Description |
|---|---|---|
| **Model sets** | | |
| gas(i) | Subset of i | Gas-fired technology |
| i | | Technology |
| m | | Month |
| n | | Node |
| nn | Alias of n | Node |
| psp(i) | Subset of i | Pump storage |
| res(i) | Subset of i | RES technology |
| reservoir(i) | Subset of i | Water reservoir |
| s | | Scenario |
| t | | (Representative) hours |
| tm(t) | Subset of t | Hours in a specific month m |
| y | | Year |
| **Model parameters** | | |
| $AF$ | % | Availability factor |
| $ARCCAP$ | $MWh_{th}$/month | Transmission capacity for the arc n → nn |
| $CAP^{existing}$ | $MW_{el}$ | Installed electricity generation capacity |
| $CC$ | $tCO_2/MWh_{th}$ | $CO_2$ emission factor per fuel consumption |
| $CHP$ | $MWh_{el}$/h | Minimum electricity generation by combined heat and power (CHP) plants to fulfil heat supply requirements |
| $CPF$ | h | Storage capacity-power factor |
| $EndLEVEL$ | $MWh_{th}$ | Stock level of gas in storage for the last period of model horizon |
| $DEMAND$ | $MWh_{el}$/h | Electricity demand |
| $DF$ | | Discount factor |
| $FLH$ | h | Full load hours of water reservoirs |
| $IC$ | €/$MW_{el}$ | Annual investment costs |
| $ICAP$ | $MWh_{th}$/month | Gas storage injection capacity |
| $ICOST$ | €/$MWh_{th}$ | Gas storage injection costs |
| $LTC$ | $MWh_{th}$/month | Long-term contract obligation for gas deliveries |
| $NPGDEM$ | $MWh_{th}$/month | Non-power sector gas demand |
| $NTC$ | $MW_{el}$ | Net transfer capacity |
| $P^{CO2}$ | €/ t $CO_2$ | $CO_2$ price |
| $PCAP$ | $MWh_{th}$/month | Available gas production capacity |
| $PCOST$ | €/$MWh_{th}$ | Marginal gas production costs |
| $PF$ | % | Hourly production factor for RES |
| $SF^{max}$ | % | Maximum shedding factor for a country |
| $StLEVEL$ | $MWh_{th}$ | Stock level of gas in storage for the first period of model horizon |
| $TCOST$ | €/$MWh_{th}$ | Marginal gas transport costs (pipelines, liquefaction and regasification processes) |
| $TOP$ | % | Take-or-pay levels |
| $VC$ | €/MWh$_{el}$ | Variable generation costs for electricity |
| $VOLA$ | €/MWh$_{el}$ | Value of lost adequacy implemented as the cost for load shedding |
| $WCAP$ | MWh$_{th}$/month | Gas storage withdrawal capacity |
| $WCOST$ | €/MWh$_{th}$ | Gas storage withdrawal costs |
| $WGV$ | MWh$_{th}$ | Working gas capacity of storage |
| $\eta$ | % | Efficiency of generation or storage technology |
| $\rho$ | % | Probability of scenario realization |
| **Model variables** | | |
| $arcflow$ | MWh$_{th}$/month | Total gas flow over the arc n → nn |
| $cap$ | MW$_{el}$ | Investments in electricity generation capacity |
| $charge$ | MWh$_{el}$/h | Electricity consumption by pumped-storage plants (PSP) |
| $flow$ | MWh$_{el}$/h | Electricity flow between nodes |
| $g$ | MWh$_{el}$/h | Electricity generation |
| $gflow$ | MWh$_{th}$/month | Gas flow volume |
| $inj$ | MWh$_{th}$/month | Gas injection volume into storage facilities |
| $level$ | MWh$_{th}$/month | Actual stock level of gas in storage |
| $pvol$ | MWh$_{th}$/month | Gas production volume |
| $shed$ | MWh$_{el}$/h | Load shedding for electricity |
| $sl$ | MWh$_{el}$ | Storage level of PSP |
| $TC$ | € | Total system costs |
| $with$ | MWh$_{th}$/month | Gas withdrawal volume from storage facilities |

**Objective function**

Objective function (11) minimises the total expected discounted capital and operating costs for both sectors:

$$\min TC = \sum_{s,y} \rho_s \cdot DF_y \cdot \begin{pmatrix} \sum_{i \backslash gas,n,t} (g_{i,n,t,y,s} \cdot VC_{i,n,t,y}) \\ + \sum_{gas,n,t} \left( g_{gas,n,t,y,s} \cdot \left[ CC_{gas} \cdot P_y^{CO2} / \eta_{gas,n,y} \right] \right) \\ + \sum_{n,t} (shed_{n,t,y,s} \cdot VOLA_n) \\ + \sum_{p,n,c} (pvol_{p,n,c,m,y,s} \cdot PCOST_{p,n}) \\ + \sum_{p,n,nn \neq n} (gflow_{p,n,nn,m,y,s} \cdot TCOST_{n,nn,t}) \\ + \sum_c (inj_{c,m,y,s} \cdot ICOST + with_{c,m,y,s} \cdot WCOST) \end{pmatrix} + \sum_{i,n,y} DF_y \cdot cap_{i,n,y} \cdot IC_i \quad (11)$$

**Integrated system constraints**

| Equation (tags: [EL] electricity sector; [G] gas sector) | Domain | Eq. |
|---|---|---|

**[EL]** Eq. (12) ensures that the market is cleared under the constraint that electricity demand in each node is satisfied at all times:

$$DEMAND_{n,t,y,s} = \sum_{i\backslash psp} g_{i,n,t,y,s} + shed_{n,t,y,s} + g_{psp,n,t,y,s} * 1/\eta_{psp,n,y} \\ - charge_{psp,n,t,y,s} + \sum_{nn}(flow_{nn,n,t,y,s} - flow_{n,nn,t,y,s}) \quad \forall n,t,y,s \quad (12)$$

**[EL]** Eq. (13) restricts hourly load-shedding activities to a share of the country-specific demand:

$$shed_{n,t,y,s} \leq DEMAND_{n,t,y,s} \cdot SF^{max} \quad \forall n,t,y,s \quad (13)$$

**[EL]** Eq.(14)-(17) define the capacity restrictions for power stations. Eq. (14) states that newly invested capacity in year *y-1* must be present in the following year *y*. Eq. (15) limits generation to the available installed capacity. Eq. (16) defines the hourly RES-feed-in. Eq. (17) considers political or technical restrictions on new investments in specific technologies (e.g. nuclear, coal):

$$cap_{i,n,y-1} \leq cap_{i,n,y} \quad \forall i,n,y \quad (14)$$

$$g_{i,n,t,y,s} \leq \left(CAP^{existing}_{i,n,y,s} + cap_{i,n,y}\right) \cdot AF_{i,n} \quad \forall n,t,y,s \quad (15)$$

$$g_{res,n,t,y,s} \leq CAP^{existing}_{i,n,y,s} \cdot PF_{res,t,n} \quad \forall RES \in I, n,t,y,s \quad (16)$$

$$cap_{i,n,y} \leq CAP^{new\ max}_{i,n,y} \quad \forall i,n,y \quad (17)$$

**[EL]** Eq. (18)–(20) describe the storage mechanism. Eq. (18) defines the maximum storage level. Eq. (19) defines the state of the storage level at the end of hour *t*. Eq. (20) defines the maximum charging capacity:

$$sl_{psp,n,t,y,s} \leq \left(CAP^{existing}_{psp,n,y} + cap_{psp,n,y}\right) \cdot CPF \quad \forall n,t,y,s \quad (18)$$

$$sl_{psp,n,t,y,s} = sl_{psp,n,t-1,y,s} - g_{PSP\,psp,n,t,y,s} + charge_{psp,n,t,y,s} \quad \forall n,t,y,s \quad (19)$$

$$charge_{psp,n,t,y,s} \leq CAP^{existing}_{psp,n,y,s} + cap_{psp,n,y} \cdot AF_{psp,n} \quad \forall n,t,y,s \quad (20)$$

**[EL]** Eq. (21) defines an annual limit to the generation by hydro reservoirs:

$$\sum_t g_{reservoir,n,t,y,s} \leq CAP^{existing}_{reservoir,n,y,s} \cdot FLH \quad \forall n,y,s \quad (21)$$

**[EL]** Eq. (22) states that gas-fired power plants are committed to country-specific CHP requirements:

$$CHP_{n,t,y} \leq \sum_{gas} g_{gas,n,t,y,s} \quad \forall n,t,y,s \quad (22)$$

**[EL]** Eq. (23) restricts cross-border electricity trading:

$$flow_{n,nn,t,y,s} \leq NTC_{n,nn,y} \quad \forall t,s \quad (23)$$

**[G]** Eq. (24) ensures that the quantity of gas imported and withdrawn from storage at each node is equal to the quantity consumed by power sectors (endogenous) and non-power sectors (exogenous) and injected into storage:

$$\sum_{p,n} pvol_{p,n,c,m,y,s} \\ = NPGDEM_{c,m,y,s} + pgdem_{c,m,y,s} + with_{c,m,y,s} \\ - inj_{c,m,y,s} \quad \forall c,m,y,s \quad (24)$$

**[G]** Eq. (25) defines capacity restrictions for gas production while eq. (26) defines those for gas transport:

$$PCAP_{p,n,y,m} - \sum_c pvol_{p,n,c,m,y,s} \geq 0 \qquad \forall p,n,m,y,s \quad (25)$$

$$ARCCAP_{n,nn,y} - arcflow_{n,nn,m,y,s} \geq 0 \qquad \forall n,nn,y$$
$$\text{where: } arcflow_{n,nn,m,y,s} = \sum_p gflow_{p,n,nn,m,y,s} \qquad \forall n,nn,m,y,s \quad (26)$$

**[G]** Eq. (27) ensures flow conservation in gas network:

$$\left[\sum_{nn\neq n} pvol_{p,n,nn,m,y,s} - \sum_{nn\neq n} gflow_{p,n,nn,m,y,s}\right] + \left[\sum_{nn\neq n} gflow_{p,nn,n,m,y,s} - \sum_{nn\neq n} pvol_{p,nn,n,m,y,s}\right] = 0 \qquad \forall p,n,m,y,s \quad (27)$$

**[G]** Eq. (28) sets a minimum amount of gas to be produced and dispatched under long-term contracts between specific nodes:

$$\sum_p pvol_{p,n,nn,m,y,s} - TOP \cdot LTC_{n,nn,m,y} \geq 0 \qquad \forall n,nn,m,y \quad (28)$$

**[G]** Eq. (29)–(32) define storage levels at the end of month $m$ and ensure intertemporal optimisation over multiple years. Eq. (33)–(35) represent constraints on storage working gas capacity, injection capacity, and withdrawal capacity:

$$level_{c,m,y,s} = level_{c,m-1,y,s} + (1-\eta) \cdot inj_{c,m,y,s} - with_{c,m,y,s} \qquad \forall c,m\backslash Jan, y,s \quad (29)$$

$$level_{c,m1,y,s} = level_{c,m12,y-1,s} + (1-\eta) \cdot inj_{c,m1,y,s} - with_{c,m1,y,s} \qquad \forall c,y,s \quad (30)$$

$$level_{c,m,y,s} = StLEVEL_c + (1-\eta) \cdot inj_{c,m,y,s} - with_{c,m,y,s} \qquad \begin{array}{c}\forall c,s \\ m=Jan, y=2020\end{array} \quad (31)$$

$$level_{c,m,y,s} \geq EndLEVEL_c \qquad \begin{array}{c}\forall c,s \\ m=Dec, y=2030\end{array} \quad (32)$$

$$WGV_{c,m,y} - level_{c,m,y,s} \geq 0 \qquad \forall c,m,y,s \quad (33)$$

$$ICAP_{c,m,y} - inj_{c,m,y,s} \geq 0 \qquad \forall c,m,y,s \quad (34)$$

$$WCAP_{c,m,y} - with_{c,m,y,s} \geq 0 \qquad \forall c,m,y,s \quad (35)$$

**[EL] [G]** Eq. (36) integrates both markets via the fuel link; gas demand of the electricity sector becomes an endogenous variable and drives gas consumption in eq. (24).

$$pgdem_{n,m,y,s} = \sum_{gas,t|t=tm} \left(g_{gas,n,t,y,s}/\eta_{gas,n,y}\right) \qquad \forall gas,t,y,s \quad (36)$$

## 3.5 Data

As already discussed, we parametrise our model primarily based on the TYNDP report from ENTSOs (2018). Section 3.5.1 details the scenarios as defined in the report while Section 3.5.2 and Section 3.5.3 present data for the electricity and gas sectors, respectively.

### 3.5.1 Introduction of scenarios

We consider scenario-dependent data for gas demand, electricity demand, installed RES capacities, fuel prices and CO2 prices. Figure 3 illustrates the three scenarios for the year 2030: distributed generation (DG), sustainable transition (ST) and the European Commission's core policy scenario (EUCO).

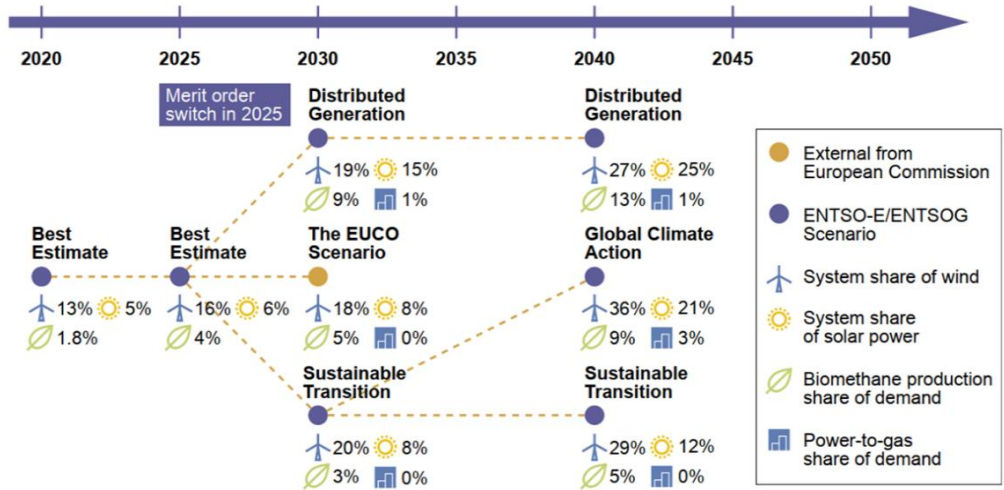

Figure 3: Scenarios from the 2018 TYNDP report. Source: ENTSOs (2018).

The DG scenario represents a decentralised development of the energy system with a focus on end-user technologies. It assumes that consumers use smart technology and dual fuel appliances (e.g. hybrid heat pumps) to switch energy sources in line with market conditions. Additionally, in this scenario, electric vehicles see their highest penetration; photovoltaics (PV) and batteries are both widely used in buildings. Relative to the other scenarios, DG is characterised by (i) the highest demand for electricity (both peak and total) and (ii) the highest amount of installed RES capacities.

The ST scenario represents a quick and economically sustainable reduction of $CO_2$ emissions achieved by replacing coal and lignite with gas in the power sector. Gas also replaces some oil usage in the transportation sector. The electrification of heat and transportation develops at a relatively slow pace. Relative to the other scenarios, ST is characterised by (i) the highest peak and total demand for natural gas and (ii) the highest price for $CO_2$ certificates (84 €/t in 2030).

EUCO, which was created using the PRIMES model and the 2016 EU Reference Scenario as a starting point, is the European Commission's core policy scenario. The scenario represents the attainment of the 2030 climate and energy targets as agreed upon by the European Council in 2014, including an energy efficiency target of 30%. Relative to the other scenarios, EUCO is characterised by (i) the highest prices for lignite (8.3 €/MWh$_{th}$ in 2030) and hard coal (15.5 €/MWh$_{th}$ in 2030) and (ii) the lowest price for $CO_2$ certificates (27 €/t in 2030).

A detailed overview of the scenario data is provided in Appendix A.[6]

### 3.5.2 Electricity sector data

The model inputs for both sectors can be roughly divided into demand, supply and infrastructure. Electricity sector demand input, thus, encompasses scenario-dependent data on country-specific load structures and annual demand projections, which stem from ENTSOs (2018).

Regarding the electricity sector supply inputs, scenario-dependent data on installed RES capacities (such as onshore wind, offshore wind and PV), fuel prices for lignite, hard coal and oil-fired power plants, and $CO_2$ prices are based on ENTSOs (2018). Prices for natural gas are endogenously derived using the integrated model. Fuel prices for nuclear power plants are based on Schröder et al. (2013). National thermal and hydro generation capacity, efficiency and decommission pathways come from Schröder et al. (2013), Gerbaulet & Lorenz (2017), European Commission (2016) and Open Power System Data (2020). Investment costs for new power stations are taken from Schröder et al. (2013). Additionally, we account for political and technical restrictions to investments in new technologies (e.g. the installation of nuclear, lignite or hard coal plants is only possible in countries without phase-out intentions). Run-of-river hydroelectricity, hydroelectric reservoirs, biomass and all above-mentioned RES are not subject to an endogenous investment decision. In order to account for country-specific CHP utilisation schemes for gas-fired units, we implement temperature-dependent must-run conditions to meet the annual production volumes of CHP plants from Eurostat (2019). The storage level of a PSP is restricted by the capacity of the upper basin. A capacity-power factor connects the installed turbine capacity with the water capacity of the upper basin; it can be understood as the full load hours of a fully charged storage plant. Generation by hydroelectric reservoirs is bounded to an annual water budget. Electricity generation from intermittent renewable capacities is not dispatchable and depends on meteorological conditions. We implement hourly feed-in profiles for onshore wind, offshore wind and PV, which are derived from the ENTSO-E Transparency Platform (2020). Given the existence of different 'wind-years', we assume that hourly feed-in profiles do not vary within our model horizon and only adjust capacity levels over time.

Within the model's geographical scope, we allow for cross-border trade. Electric power transmission between nodes is restricted by net transfer capacities, which are from ENTSOs (2018). Intranational imitations on electricity flows are neglected in this study. Load-shedding activities, which are driven by a scarcity of power plant capacities, are penalised by the value of lack of adequacy (VoLA), which is determined for each European country individually by Cambridge Economic Policy Associates (2018). In order to avoid an unreasonable 'trade' of shedding activities, we implement a maximum shedding factor that is assumed to limit hourly load shedding to 20% of respective hourly demand in a node.

---

[6] A discussion of the assumptions for each scenario and the background methodology of the TYNDP report can be found on the following website: tyndp.entsoe.eu/tyndp2018/scenario-report/

### 3.5.3 Gas sector data

Scenario-dependent data on gas demand projections for European countries are based on scenarios from the ENTSO-G (2018b). The annual gas demand levels are broken down to a monthly structure for each node. Monthly demand profiles are calculated based on historical average monthly gas consumption data from Eurostat (2019).

We also use the ENTSO-G (2018b) for data on gas supply potential. We consider long-term contracts on an annual level, in line with Neumann et al. (2015), to realistically represent gas market fundamentals. In particular, we use information on contracting parties, annual contracted gas volume and contract expiration dates.[7] These data are used in the model as an exogenous constraint specifying the minimum bound on a trade variable between respective nodes. This constrains diversification of supplies from importing countries that would not have been captured if the long-term obligations had been omitted.

Data on the existing gas pipeline infrastructure are from the ENTSO-G (2018a) capacity map. Data for LNG infrastructure are based on GIE (2019) and GIIGNL (2016). Data about national storage capacities are based on GIE (2019). All storage data are aggregated on the node level (i.e. each region has one representative storage node). Strategic storage requirements are based on an European Comission (2015) report.[8] Our model incorporates exogenous gas infrastructure capacity expansions. The structure of the system's development is harmonized with the information from the ENTSO-G (2018b). Only units with final investment decision status are included in the dataset.

We used numerous public information portals and academic studies to parametrise the cost structure of gas production (Chyong & Hobbs, 2014), transmission (Chyong & Hobbs, 2014; ACER, 2018; Rogers, 2018; Fodstad et al., 2016) and storage (European Comission (2015), 2015). See Riepin & Müsgens (2019) for details on the cost structure and necessary assumptions for the gas model.

## 4 RESULTS

This section is organised in the following way. We begin with a brief overview of the solutions from a deterministic investment problem of an integrated electricity and gas system in Section 4.1. We then detail the costs of ignoring uncertainty in a composite mode in Section 4.2. Finally, in Section 4.3, we focus on the effects of single parametric uncertainty.

---

[7] As information about take-or-pay levels is not disclosed, we assume a level of 70%.
[8] Thus, country-specific shares of storage capacities, which are booked for strategic storage, are exogenously fixed and excluded from the model's decision space.

## 4.1 Deterministic solutions

We begin with a classical scenario analysis by formulating a deterministic cost-minimisation investment problem as defined in equation (1) and incorporating different inputs from three ENTSOs scenarios and an EVP.

Note that model's investment decision space is limited to thermal power plants (open-cycle gas turbines [OCGT], combined-cycle gas-turbines [CCGT], and lignite and hard coal power plants) and PSPs. Capacity installations of wind, PV, biomass, hydroelectric reservoirs and run-of-river technologies, based on data from the 2018 TYNDP, are implemented exogenously.

Figure 4 breaks down the investment mix by scenario. Evidently, the different assumptions (input data) across the four scenarios result in different solutions (optimal technology mix) to the deterministic cost-minimisation problems.

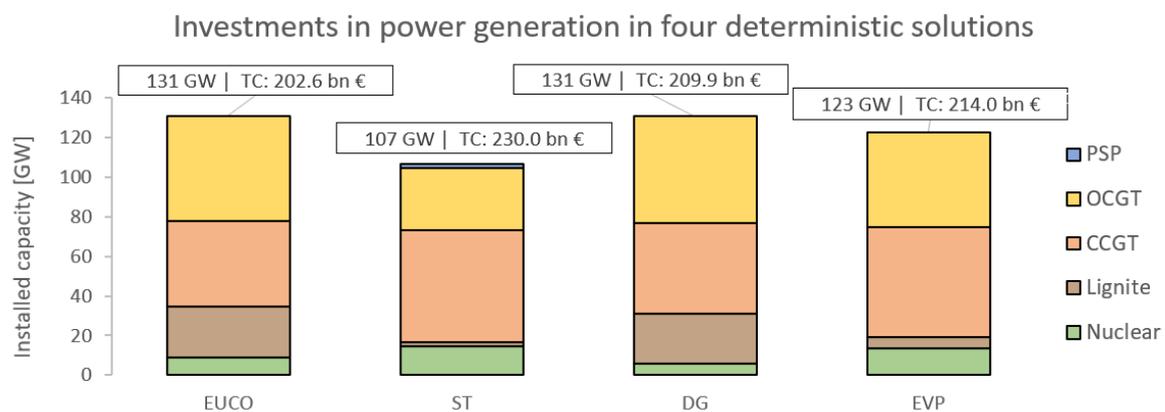

Figure 4: Investments in power generation capacities [GW] and total system costs [bn €$_{2020}$] for three deterministic scenarios and an expected value problem.

The EUCO and DG scenarios are characterised by fairly similar investment mixes. A considerable amount of lignite capacities (ca. 25 GW in each) in these investment mixes can be explained by the middle-term 2020–2030 modelling horizon (i.e. we do not account for further climate policy actions after 2030). In particular, $CO_2$-intensive capacities that are built with such foresight may end up stranded later on.[9] In the ST scenario, the optimal investment decisions are based on, among other influential factors, an expectation of a high $CO_2$ price. Thus, we observe the highest degrees of investment in nuclear (14.5 GW) and CCGT (56.9 GW) technologies due to the high efficiency of $CO_2$ emissions per MWh$_{el}$. Interestingly, in the EVP scenario, capacities invested in each technology are not close to a mere arithmetical mean among the scenarios. The capacities of CCGT and nuclear technologies are similar to those in the ST scenario, though just a minor amount of lignite is kept in the optimal investment mix.

Another thing that stands out in Figure 4 is the difference in total system costs across individual deterministic solutions; EUCO has the lowest aggregated and discounted investment costs (€ 202.6 bn) followed by DG (€ 209.9 bn), EVP (€ 214.0 bn) and, the most expensive, ST (€ 230.0 bn). These cost differences are, again, driven by scenario-specific assumptions. The high total system costs in the ST scenario are driven by the high variable costs of power generation, which, in part, are driven by the scenario's relatively high $CO_2$ price.

---

[9] Löffler et al. (2019) illustrate a stranded assets problem.

As already discussed, an alternative to such a scenario analysis is to explicitly encode uncertainty as a part of the model by setting a probability distribution for uncertain parameters, as shown in equations (2) and (3). Comparing the two approaches—ignoring uncertainty versus explicitly modelling uncertainty—is crucial for us understand the degree to which modelling uncertainty affects decision quality.

## 4.2 Effects of ignoring uncertainty: All parameters

In this section, we continue to solve the stochastic problem defined in equations (2) and (3) by using a vector of uncertain data that includes all five uncertain parameters. The optimal first-stage decision for this problem includes a mix of technologies that hedges against a composite uncertainty (as the stochastic model, by definition, minimises the expected costs by accounting for all possible scenarios).

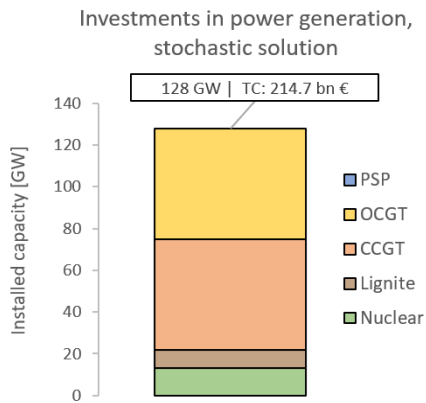
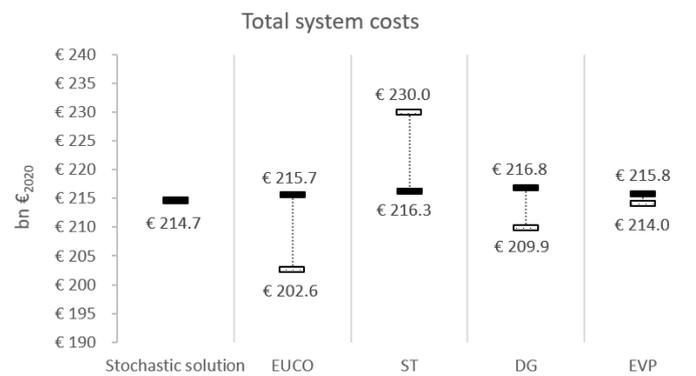

Figure 5: Investments in power generation capacities [GW] for the stochastic solution.

Figure 6: Total expected system costs for the SS and four EEV (markers with a solid fill) and total system costs for four DS (markers with a pattern fill).

Figure 6 highlights how the scenario-specific deterministic first-stage investment decisions perform in a stochastic setting ($EEV_n$, markers with a solid fill). Interestingly, the optimal investment mix for the stochastic problem is not optimal for any individual scenario despite the fact that the stochastic solution has the lowest *expected* costs overall.

The difference between the $EEV_n$ and the $SS$ is $ECIU_n$ (as shown in equation (10)). Table 1 depicts the results for the ECIU calculation. Note that ECIU cannot be negative, as recognising the correct probabilities cannot worsen the expected cost (van der Weijde & Hobbs, 2012).

Evidently, the ECIU varies by scenario. The ECIU values range from € 1,005.2 M (EUCO scenario) to € 2,131.3 M (DG scenario). For comparison, this accounts for 10.1–21.3% of all power generation investment costs in the stochastic solution.[10]

Table 1: The expected costs of ignoring uncertainty: A composite mode (considering all uncertain parameters)

|  | EUCO | ST | DG | EVP |
|---|---|---|---|---|
| ECIU [€$_{2020}$][1] | € 1,005.2 M | € 1,620.5 M | € 2,131.3 M | € 1,125.4 M |
| ECIU [% of investment costs] | 10.1% | 16.2% | 21.3% | 11.3% |

[1] Costs are computed for three representative years: 2020, 2025 and 2030.

In contrast to intuition, the ECIU in the EUCO scenario is lower than that in the EVP scenario. This is likely due to load-shedding activities, which are driven by the allocation of scenario-specific maximum residual load levels. A closer inspection of ENTSOs data for electricity demand reveals there are several country subsets (i.e. nodes in our model) that face maximum residual load levels in each of the three scenarios. Hence, no investment plan based on a deterministic scenario is dominant with regard to minimising load-shedding activities in the stochastic problem. A detailed analysis of this issue is provided in Appendix B.

### 4.3 Effects of disregarding uncertainty: Isolated parameters

This section addresses isolated uncertainties. For this, we set a vector of uncertain outcomes in a stochastic problem to include only one unknown parameter, allowing us to compute the isolated effects of parametric uncertainty (see Section 3.3 for details on scenario composition). Table 2 presents the results which are backed up with a detailed analysis below.

Table 2: Expected costs of ignoring uncertainty: Isolated parametric uncertainty

| Isolated parameter | EUCO | ST | DG | EVP |
|---|---|---|---|---|
| Electricity demand | € 577.3 M | € 695.8 M | € 1,112.6 M | € 674.7 M |
| Installed RES capacity | € 89.9 M | € 83.0 M | € 389.5 M | € 104.5 M |
| Gas demand[1] | € 50.2 M | € 45.9 M | € 3.0 M | € 0.4 M |
| Fuel price[2] | € 245.8 M | € 9.3 M | € 51.0 M | € 1.0 M |
| $CO_2$ price | € 864.2 M | € 661.4 M | € 11.5 M | € 41.9 M |

Costs are computed for three representative years: 2020, 2025 and 2030. [1] 'Gas demand' implies uncertainty in non-power sector gas demand; [2] 'fuel price' reflects uncertainty in lignite, hard coal and oil prices.

#### 4.3.1 Electricity demand and installed RES capacity

As shown in Table 2, ECIU for electricity demand is significant across all four scenarios. ECIU is the highest, among all parameters, for electricity demand in three of the four scenarios. This is likely because investment decisions are sensitive to peak demand levels. Underestimating this effect leads to under-investment that cause higher load-shedding activities while overestimating it leads to over-investments and, in turn, higher investment costs.

---
[10] The costs contain discounted investment payments for the model's three representative years.

Under the condition of uncertain electricity demand, investment decisions derived using deterministic scenarios, compared to the stochastic problem, lead to higher shedding costs in the EEV problems (see Table 5 in Appendix B). This is due to the fact that scenario-specific demand developments differ between European countries. Thus, all deterministic capacity expansion plans show under-investment in some nodes and, therefore, increased shedding activities. The DG scenario, for example, has the highest electricity demand (see scenario data in Appendix A), which results in an investment level higher than that of the stochastic solution. This, in turn, leads to lower load-shedding activities relative to other scenarios. However, certain nodes' demand levels reach their maximum in other scenarios (e.g. in the UK node, maximum demand is observed in the EUCO scenario; thus, constraining the stochastic problem with the investment mix from the DG scenario results in increased shedding activities in the UK node).

As with electricity demand, electricity generation from RES capacities eventually affects the residual load. Hence, the mechanism how uncertainty in RES installed capacities affects the investment decisions is similar to that in electricity demand (i.e. a system planner optimises a capacity mix that varies by scenario- and country-specific residual load levels). However, when analysing uncertain development in RES capacity, the ECIUs are lower than those for electricity demand. This is likely due to the influence of parametric uncertainty on the residual load.

Figure 7 shows boxplots for hourly residual load levels for each scenario in 2030 after varying after varying electricity demand and RES capacity levels. The necessary assumption to depict residual load levels for electricity demand variation is to fix the RES feed-in to one of four scenarios (and vice versa for RES feed-in variation). We use the EUCO scenario for both illustrative examples in Figure 7 but the impact on the residual load is within the same magnitude for the other scenarios. Residual load boxplots for all scenarios are in Appendix B.

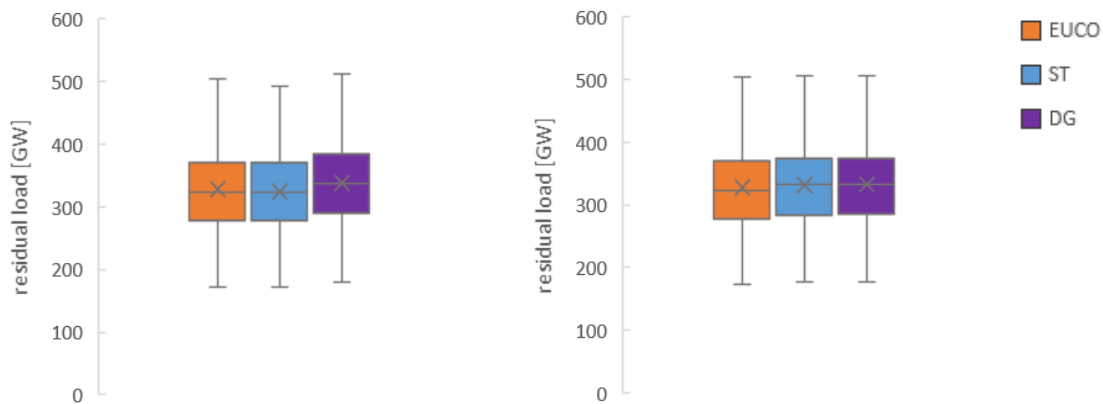

Figure 7: Comparing residual load after varying demand (left) and RES capacity (right) following ENTSO-E scenarios for 2030.

It is clear that the difference in maximum residual level between scenario branches is significantly higher when implementing demand variation. The maximum residual load level ranges within the three branches from 491.2 GW to 512.0 GW when electricity demand is varied but only from 503.6 GW to 506.2 GW when RES capacity is varied.

### 4.3.2 Gas demand, fuel price and $CO_2$ price

As shown in Table 2, ECIU for gas demand is at a negligible level across all four scenarios. ECIU for fuel price is negligible in all but one scenario (EUCO). The effect of $CO_2$ price uncertainty strongly varies by scenario from high to negligible. We back up these observations with a detailed analysis of the impact that parametric uncertainties have on the integrated system.

Our analysis is supported by Figure 8, which comprises convex hull plots depicting variable costs for four technologies: lignite, hard coal, CCGT and OCGT. Nuclear and biomass are excluded because their costs do not differ across ENTSOs scenarios. The abscissa represents years. The ordinate represents variable costs in €/MWh. Each column represents an uncertainty driver—gas demand, fuel price and $CO_2$ price—while each row represents a scenario path for the other four known parameters. Each hull is built around nine dots—three dots per year. The three dots represent variable costs for each technology evaluated at gas, fuel, and $CO_2$ prices observed in each of the three branches in a relevant stochastic model. Thus, convex hull width conveys ranges of variable costs per technology caused by parametric uncertainty. These plots are useful because they visualize how parametric uncertainty affects variable costs of power generation and provide an intuition for the resulting ECIU measures.[11]

In the first column of Figure 8, convex hull width depicts the sensitivity of variable costs to gas demand uncertainty. The impact of gas demand uncertainty on the variable costs of CCGT and OCGT is small. Predictably, the impact on lignite and hard coal is null. We conclude that the differences in gas prices driven by the uncertainty of gas demand do not significantly change installed capacities of electricity generation technologies. This result is driven by the substantial European gas infrastructure, which is sufficiently capable of meeting the range of future gas demand scenarios defined in TYNDP. The negligible ECIU measures for gas demand uncertainty reflect this observation.[12]

The second column of Figure 8 depicts the sensitivity of variable costs to the uncertainty of lignite and hard coal prices. It is notable that the intersection of lignite and CCGT significantly varies by underlying fuel price assumption. For example, with uncertain fuel prices and the high $CO_2$ price in the ST scenario, generation with CCGT becomes cheaper than that with a lignite plant in all branches of the stochastic model by 2025. Isolating fuel price uncertainty reveals that the cost structures of the ST, DG and EVP scenarios ensure that there are null or minor investments in lignite power plants. Thus, uncertainty in lignite prices has a negligible effect. In the EUCO scenario, with a low $CO_2$ price, lignite remains cheaper than CGGT all over the modelling horizon in some branches of a stochastic problem. Consequently, the optimal solution contains a moderate amount of lignite capacity in the investment mix. In this case, lignite price uncertainty has a notable effect, which is reflected by the ECIU measure.

---

[11] More detail on constructing convex hull plots can be found on the following link:
github.com/Irieo/IntEG/blob/master/HullPlots/Hulls.ipynb

[12] Riepin et al. (2018) illustrate the reallocation of gas-fired technologies as the effect of gas demand uncertainty on electricity sector investments. However, they also point out that the value of a stochastic solution (the ECIU) is small. Similar results are revealed by Fodstad et al. (2016), who study the effect of gas demand uncertainty on investments in gas sector infrastructure. They find structurally distinct infrastructure solutions in stochastic and deterministic models but report a negligible value of a stochastic solution.

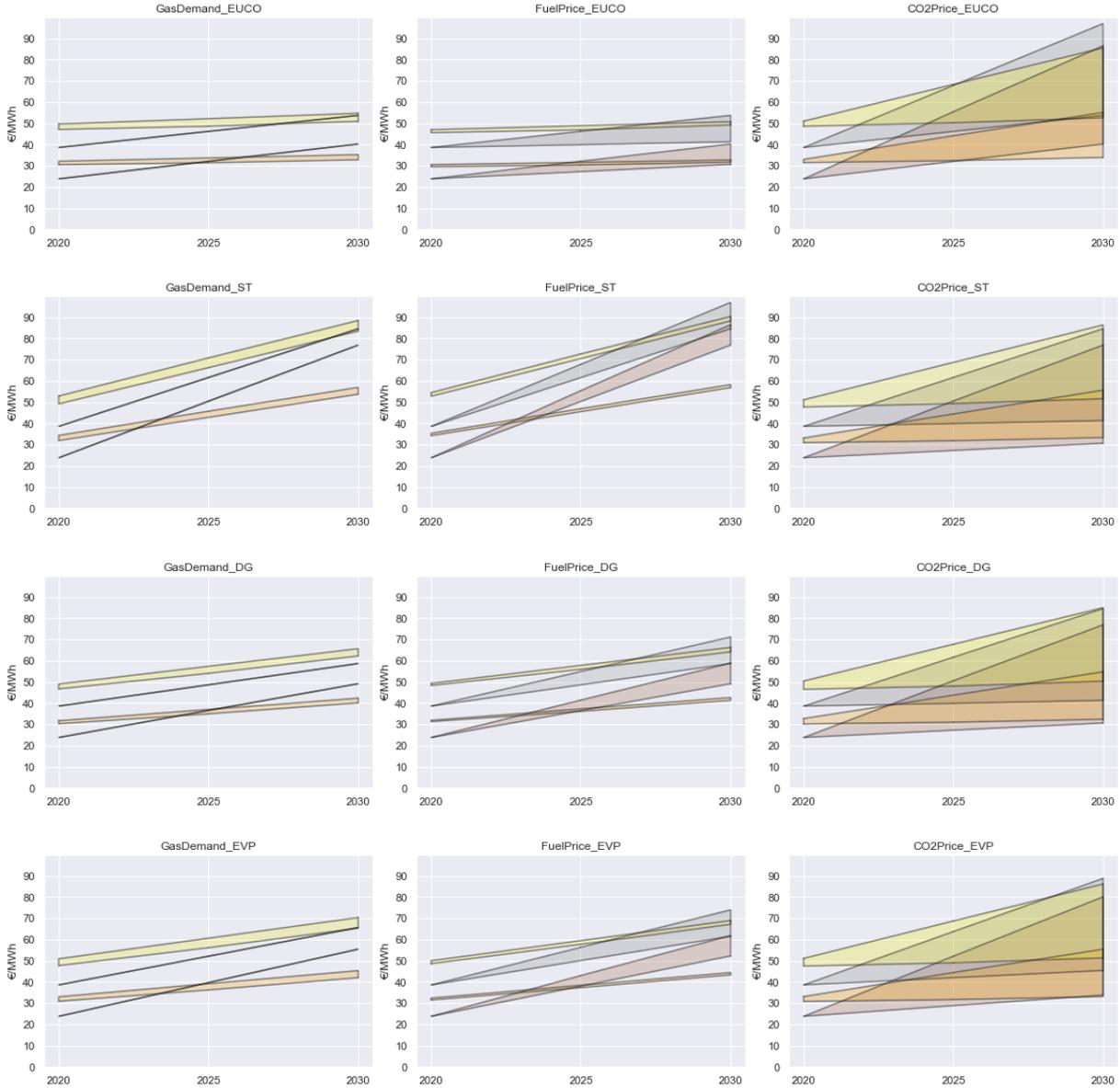

Figure 8: Variation of variable costs for four technologies due to uncertainty regarding gas demand (first column), fuel price (second column) and $CO_2$ price (third column).

Colour codes: lignite [brown]; hard coal [grey]; CCGT [orange]; OCGT [yellow].

The third column of Figure 8 depicts the sensitivity of variable costs to the uncertainty of $CO_2$ prices. The hull widths (ranges of variable costs) are higher than in the other two parametric uncertainties. In order to explain values for $CO_2$ price uncertainty in Table 2, it is important to note that hull width illustrates the potential for high costs of ignoring uncertainty. This is driven by the stochastic problem hedging against the large differences in variable costs established by the extreme $CO_2$ price values (the 'higher' and 'lower' hull edges). The deterministic solution, by definition, disregards this variation and seeks to optimise investments for a particular $CO_2$ price realisation. The width of hulls (i.e. variety of cost structures) highlights this mismatch. This is illustrated by the results in the EUCO and ST scenarios (characterised by the high and low $CO_2$ prices, respectively), resulting in considerable ECIU measures. The optimal capacity mix for deterministic solutions in these two scenarios significantly differs from the optimal mix for the stochastic problem, which accounts for a full range of scenarios. Therefore, evaluating the performance of these deterministic solutions in a stochastic setting results in considerably higher expected system costs.

# 5   CONCLUSIONS

This paper investigates the trade-off between complexity from integrated optimisation of gas and electricity systems and parametric uncertainty. To do this, we use a combination of deterministic and stochastic optimisation approaches. The methodological contributions of combining integrated and stochastic optimisation problems and comparing the isolated effects of parametric uncertainty shall be of interest for energy modellers. We parametrise our model and derived scenarios from data in the 2018 TYNDP report published by ENTSOs. We believe that our findings are of interest to industry experts and stakeholders with an empirical interest in the European energy system. In order to enhance the transparency and reproducibility of our results, we have published the data and source codes for the entire research project online. Beyond these broad methodological and empirical contributions, we recognize five key take-aways.

First, the expected costs of ignoring uncertainty can constitute a significant share (up to 20%) of cumulated costs from investments in power generation capacity. Although our results are obtained under restrictive assumptions and do not reflect all real-world planning complexities, this number is impressive given that it is obtained just by adjusting the planning approach.

Second, concerning the underestimation of required capacity and the resulting load-shedding activities, our results suggest that no capacity expansion plan based on a deterministic scenario is dominant in terms of minimising load-shedding activities. Energy modellers should recognise that, in a framework of European TYNDP data, choosing the scenario with the highest aggregated demand does not result in the absence of load shedding on account of country-specific load peak variation.

Third, our analysis of isolated parametric uncertainties revealed that the effect of ignoring electricity demand uncertainty is significant across all four scenarios in our modelling scope. For installed RES capacities, the effect is notably low. The magnitude of the effect for these parametric uncertainties is mainly driven by residual load peak variation. This observation should be seen in light of the ten-year modelling horizon we included; however, it strongly suggests incorporating electricity demand uncertainty into research items that focus on the expansion plans of power generation in the context of increasing capacities of renewable energy sources.

Fourth, the expected costs of disregarding gas demand uncertainty are low across all considered scenarios. The take-away for energy modellers is two-fold. One, our findings suggest that there is a small value for incorporating stochastic optimisation under gas demand uncertainty into large-scale integrated electricity and gas system models. Two, when focusing on the impacts of long-term uncertainty in single-sector models for electricity and gas, the effect of capturing interactions between the two sectors seems limited; this could justify the choice of future modellers to focus resources on a single sector when analysing uncertainty. It is worth noting that this finding should be seen in the light of our modelling scope and TYNDP scenario data for gas demand uncertainty.

Fifth, the effect of $CO_2$ price uncertainty strongly varies by the scenario path of other parameters. In the context of the 2018 TYNDP scenarios, the effect is negligible when the DG scenario realises. In contrast, ignoring $CO_2$ price uncertainty leads to expensive decisions if the EUCO or ST scenarios play out, as investments made in anticipation of these scenarios poorly match with system needs after uncertainty is revealed. Our concluding suggestion for the energy modelling community focused on long-term system planning problems is to consider including $CO_2$ price uncertainty into model formulations—at least during test runs. Investigating the interactions between assumed future $CO_2$ price ranges (or, alternatively, emission caps) and scenario assumptions for the other parameters can bring significant benefits over naïve modelling practices.

Further research on this subject is necessary. In terms of methodology, the major assumption of the stochastic optimisation framework used in this paper is that there are two stages: before a parametric uncertainty is realised and after a parametric uncertainty is realised. However, this is a simplification of reality; at no point are energy market players exposed to complete and perfect information. Thus, further research and advances in computational capabilities may address questions on multi-stage modelling approaches raised by this study. More work must be done to investigate the impact of uncertainties that were not discussed in this paper, such as gas sector supply. Furthermore, as our model was formulated with linear programming, developments in demand for electricity and natural gas are modelled as completely price-inelastic. This is a simplification, especially in the long run. Researching this topic while considering demand elasticity could provide useful new insights.

# 6   DATA AVAILABILITY

Datasets related to this article and a source code for the entire project are available in the public GitHub repository: github.com/Irieo/IntEG. The code reproduces the benchmarks from the paper.

# APPENDIX A: SCENARIO DATA

## GAS DEMAND

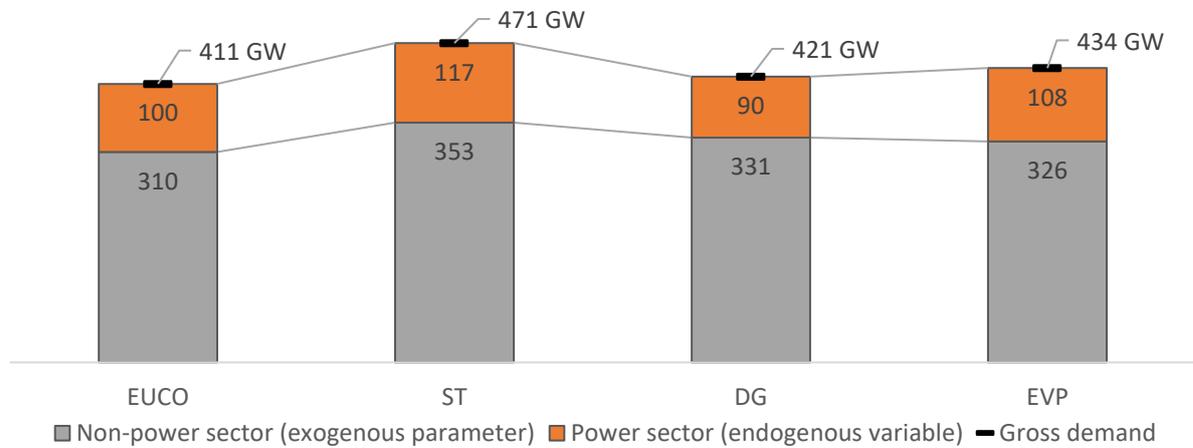

Figure 9: Breakdown of annual European gas demand in 2030, GW. Projections for non-power sector gas demand (grey) are based on the 2018 TYNDP report. Power sector gas demand (orange) is determined endogenously in our model.

## ELECTRICITY DEMAND

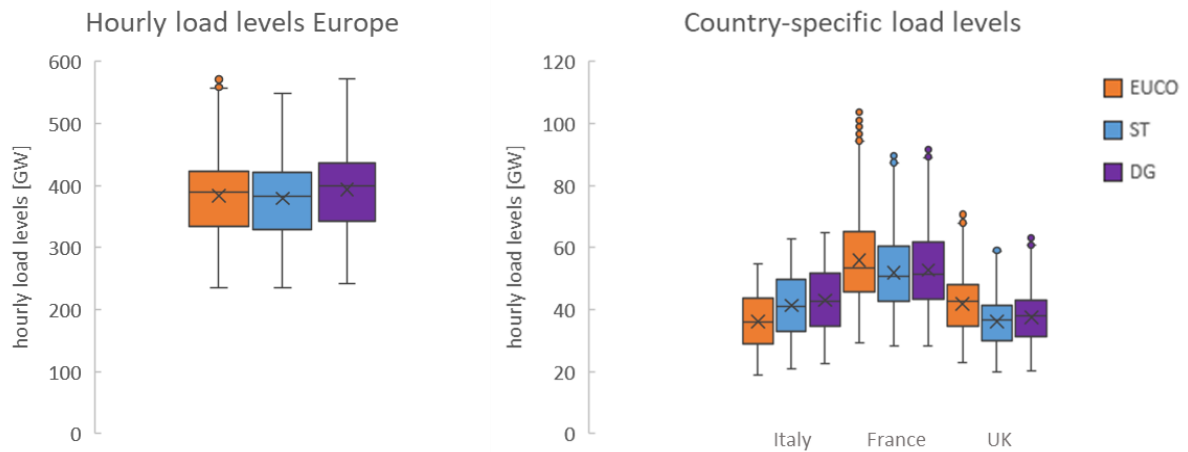

Figure 10: Projections for European electricity demand in 2030 (left); variation of load levels for three selected countries in 2030 (right).

## INSTALLED RES CAPACITY

Table 3: Projections for installed RES capacities, GW.

|  | 2015 Historic | 2020 Best estimate | 2025 EUCO | 2025 ST | 2025 DG | 2030 EUCO | 2030 ST | 2030 DG |
|---|---|---|---|---|---|---|---|---|
| Onshore wind | 126.7 | 161.8 | 201.3 | 193.5 | 193.5 | 240.7 | 225.1 | 225.1 |
| Offshore wind | 11.7 | 28.1 | 34.3 | 46.1 | 46.1 | 40.5 | 64.2 | 64.0 |
| PV | 64.0 | 98.2 | 138.0 | 186.4 | 191.6 | 299.7 | 230.5 | 238.3 |

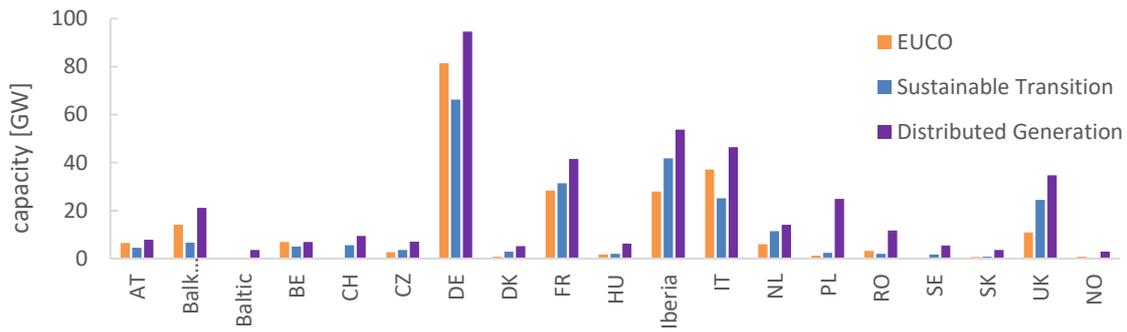

Figure 11: Projections for installed PV capacities in 2030.

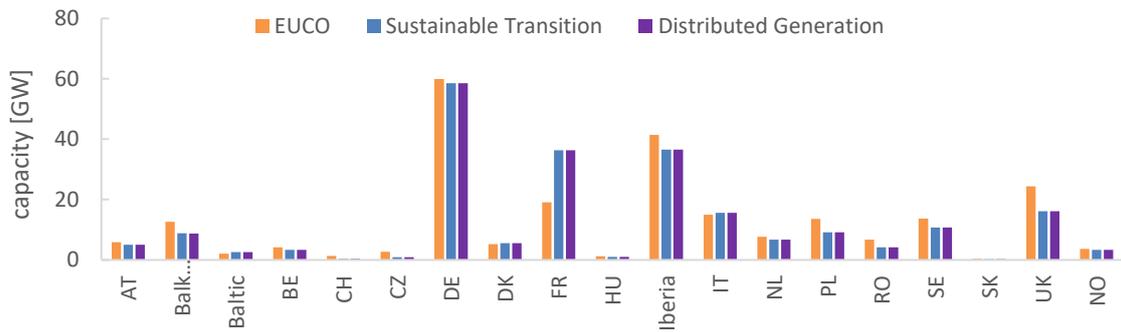

Figure 12: Projections for installed onshore wind capacities in 2030.

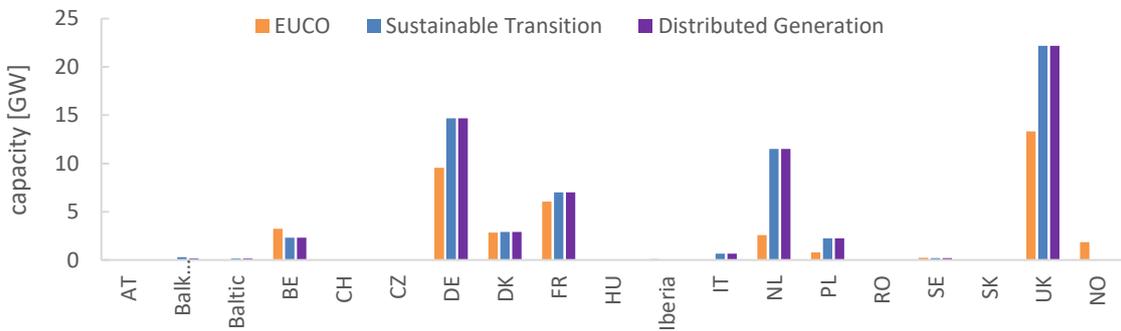

Figure 13: Projections for installed offshore wind capacities in 2030.

## FUEL AND $CO_2$ PRICES

Table 4: Fuel and $CO_2$ prices. Monetary values are in €$_{2015}$.

|  |  | **2015** Historic | **2020** Best estimate | **2025** EUCO | **2025** ST | **2025** DG | **2030** EUCO | **2030** ST | **2030** DG |
|---|---|---|---|---|---|---|---|---|---|
| $CO_2$ price | €/ton $CO_2$ | 8.00 | 18.00 | 22.50 | 51.15 | 34.00 | 27.00 | 84.30 | 50.00 |
| Nuclear | €/MWh$_{th}$ | 1.69 | 1.69 | 1.69 | 1.69 | 1.69 | 1.69 | 1.69 | 1.69 |
| Lignite | €/MWh$_{th}$ | 4.80 | 3.96 | 6.12 | 3.96 | 3.96 | 8.27 | 3.96 | 3.96 |
| Hard coal | €/MWh$_{th}$ | 8.17 | 8.27 | 11.87 | 8.99 | 8.99 | 15.47 | 9.71 | 9.71 |
| Oil | €/MWh$_{th}$ | 23.83 | 55.76 | 64.75 | 67.09 | 67.09 | 73.74 | 78.42 | 78.42 |
| Biomass | €/MWh$_{th}$ | 8.10 | 9.0 | 9.90 | 9.90 | 9.90 | 10.80 | 10.80 | 10.80 |

# APPENDIX B: SUPPLEMENTARY MATERIAL

## SUPPLEMENTARY MATERIAL FOR SECTION 4.2

No investment plan based on a deterministic scenario is dominant with regard to minimising load-shedding activities in the stochastic problem. Each stochastic problem with fixed investments has higher shedding costs than the stochastic problem (all delta terms with regard to shedding costs in Table 5 are positive).

An intuitive assumption is that extended shedding costs are caused by a lack of investment, implying that savings in investment cost will take place. This assumption is supported by Table 5, which depicts aggregated investment costs. In three out of four scenarios, the investments based on deterministic scenarios are less costly than those based on the stochastic problem (IV). However, the saved investment costs are not able to equalize increased shedding costs (II and IV). In the ST scenario, savings in investment costs are negative (i.e. investments cost more than in the stochastic solution).

The difference between increasing shedding costs and decreasing investment costs is lowest with an investment plan determined by the EUCO scenario assumptions—this explains the lowest ECIU value among the four scenarios (see discussion in Section 4.2).

Table 5: Breakdown of shedding and investment costs.

|  |  | Stochastic problem (SP) | Stochastic problem with fixed investments (EEV) | | | |
|---|---|---|---|---|---|---|
|  |  |  | EUCO | ST | DG | EVP |
| I: Shedding costs | bn €$_{2020}$ | 0.48 | 0.92 | 2.76 | 2.13 | 1.69 |
| II: Delta shedding costs (EEV – SS) | bn €$_{2020}$ | - | 0.44 | 2.28 | 1.65 | 1.21 |
| III: Investment costs | bn €$_{2020}$ | 10.00 | 9.92 | 10.32 | 9.20 | 9.57 |
| IV: Delta investment costs (EEV – SS) | bn €$_{2020}$ | - | -0.08 | 0.32 | -0.80 | -0.43 |

SUPPLEMENTARY MATERIAL FOR SECTION 4.3.1

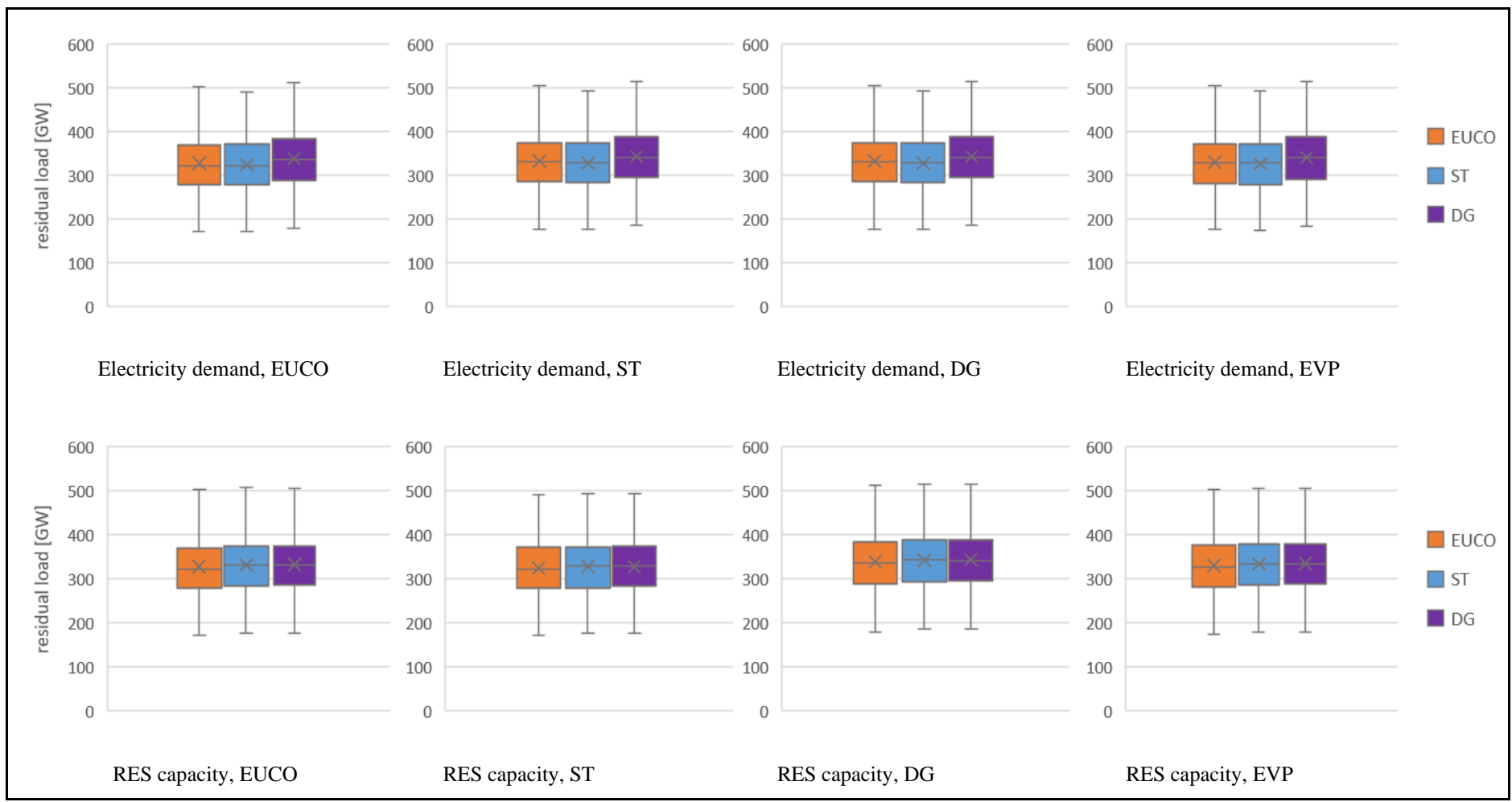

Figure 14: Boxplots of European residual load variations due to electricity demand uncertainty (upper row) and RES capacity uncertainty (lower row) per the ENTSO-E scenario for 2030.

# APPENDIX C: GEOGRAPHICAL SCOPE

Below, we list the countries and regions included in our model's geographical scope.

Note that the model structure consists of a network of nodes. A node represents one country or a group of several countries from one region. The geographical scope covers EU member states, European countries that are not part of the EU and major non-European gas exporters. We elicit node names as defined by the two-letter ISO 3166 international standard. In cases where countries are represented by a compound region with aggregated data, we set the node name in line with the name of the region.

| Country | Node | Electricity sector components | Gas sector components |
|---|---|---|---|
| *Nodes representing EU member states*[13] | | | |
| Austria | (AT) | yes | yes |
| Belgium | (BE) | yes | yes |
| Czech Republic | (CZ) | yes | yes |
| Denmark | (DK) | yes | yes |
| France | (FR) | yes | yes |
| Germany | (DE) | yes | yes |
| Hungary | (HU) | yes | yes |
| Italy | (IT) | yes | yes |
| Netherlands | (NL) | yes | yes |
| Poland | (PL) | yes | yes |
| Romania | (RO) | yes | yes |
| Slovakia | (SK) | yes | yes |
| Sweden | (SE) | yes | yes |
| *Nodes representing EU regions* | | | |
| Bulgaria | (Balkans) | yes (aggregated) | yes (aggregated) |
| Greece | (Balkans) | yes (aggregated) | yes (aggregated) |
| Croatia | (Balkans) | yes (aggregated) | yes (aggregated) |
| Slovenia | (Balkans) | yes (aggregated) | yes (aggregated) |
| Serbia[14] | (Balkans) | yes (aggregated) | yes (aggregated) |
| Estonia | (Baltics) | yes (aggregated) | yes (aggregated) |
| Latvia | (Baltics) | yes (aggregated) | yes (aggregated) |
| Lithuania | (Baltics) | yes (aggregated) | yes (aggregated) |
| Spain | (Iberia) | yes (aggregated) | yes (aggregated) |
| Portugal | (Iberia) | yes (aggregated) | yes (aggregated) |
| *Nodes representing European countries that are not EU member states* | | | |

---

[13] EU member states not represented in the model are Cyprus, Finland, Luxembourg and Malta.

[14] As of June 2020, Serbia is not an EU member state; however, the country is included as a part of the Balkans node due to regional importance.

| Switzerland | (CH) | yes | yes |
| United Kingdom | (UK) | yes | yes |
| Norway | (NO) | yes | yes (supply side only) |
| *Nodes representing non-European countries that are major gas exporters to the EU* | | | |
| Algeria | (DZ) | no | yes (supply side only) |
| Libya | (LY) | no | yes (supply side only) |
| Nigeria | (NG) | no | yes (supply side only) |
| Qatar | (QA) | no | yes (supply side only) |
| Russia | (RU) | no | yes (supply side only) |
| United States | (US) | no | yes (supply side only) |